\documentclass{aip-cp}

\usepackage[numbers]{natbib}
\usepackage{rotating}
\usepackage{graphicx}
\usepackage{xspace}

\newcommand{\pb}{{Pb--Pb}\xspace}
\newcommand{\pPb}{{p--Pb}\xspace}
\newcommand{\RunOne}{Run~1\xspace}
\newcommand{\RunTwo}{Run~2\xspace}

\newcommand{\LSOne}{LS1\xspace}
\newcommand{\LSTwo}{LS2\xspace}

\newcommand{\pt}{\ensuremath{p_{\mathrm{T}}}\xspace}
\newcommand{\nbinv}{\ensuremath{\mathrm{~nb}^{-1}}\xspace}
\newcommand{\jpsi}{\ensuremath{\mathrm{J}/\psi}\xspace}

\begin{document}

\title{ALICE Upgrades: Plans and Potentials}

\author[]{R. Tieulent$^1$ on behalf of the ALICE Collaboration} 

\affil[aff1]{Universit\'e de Lyon, Universit\'e Lyon 1, CNRS/IN2P3, IPN-Lyon, Villeurbanne, France}

\maketitle

\begin{abstract}
The ALICE collaboration consolidated and completed the installation of current detectors during \LSOne with the aim to accumulate 1~\nbinv of \pb collisions during \RunTwo corresponding to about 10 times the \RunOne integrated luminosity.
In parallel, the ALICE experiment has a rich detector upgrade programme scheduled during the second LHC long shutdown (LS2, 2018--2019) in order to fully exploit the LHC Runs 3 and 4.
The main objectives of this programme are: (a) improving the tracking precision and (b) enabling the read-out of all \pb interactions at a rate of up to 50~kHz, with the goal to record an integrated luminosity of 10~\nbinv after \LSTwo in minimum-bias trigger mode. This sample would represent an increase by a factor of one hundred with respect to the minimum-bias sample expected during \RunTwo.
The implementation of this upgrade programme, foreseen in \LSTwo, includes: 
a new low-material Inner Tracking System at central rapidity with a forward rapidity extension (MFT) to add vertexing capabilities to the current Muon Spectrometer; 
the replacement of the Time Projection Chamber (TPC) wire chambers with gas electron multiplier (GEM) readout;
a new readout electronics for most of the detectors and an updated trigger system;  a new set of forward trigger detectors and a new integrated online--offline system.
High-precision measurements of open heavy-flavour, quarkonia and low-mass dilepton production are the major physics goals of the ALICE upgrade programme.
Thanks to the new inner tracking system, the measurements of the nuclear modification factor
and elliptic flow of several species of charm and beauty mesons and baryons will be extended to zero or close to zero \pt.

At forward rapidity, the new silicon based tracker (MFT) will allow prompt J/psi to be separated from displaced J/psi from B decays down to zero pT as well as single muons to be separated from charm and beauty hadron decays.
\end{abstract}

\section{INTRODUCTION}

\begin{figure}[h]
 \centerline{\includegraphics[width=0.9\textwidth]{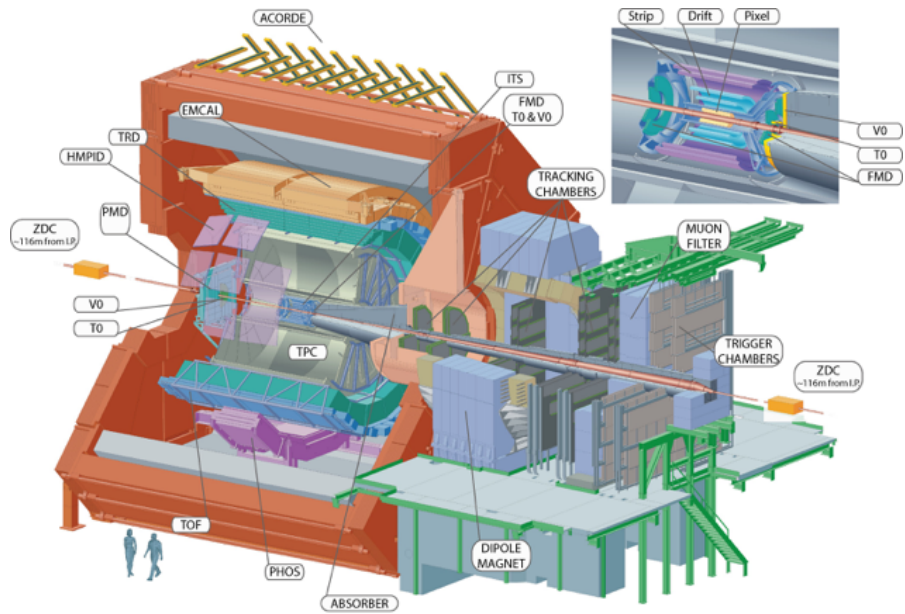}}
  \caption{The ALICE experiment setup during LHC Run 1.}
  \label{fig:ALICE}
\end{figure}

ALICE is the experiment totally devoted  at LHC to the study of the Quark Gluon Plasma which is a deconfined state of matter produced in heavy-ion collisions.
The LHC gives the possibility to study the QGP at almost zero baryonic density, close to the path followed by the early Universe while cooling down.
At LHC, the temperature of the medium produced in \pb collisions is much higher than the transition temperature which makes the LHC the best facility to study QGP properties.
ALICE performs these studies in a broad \pt range using several probes including heavy-flavour, quarkonia, light hadrons, jets, photons and more.
The ALICE detectors are separated into two main components (see Figure~\ref{fig:ALICE}), the central barrel covering the mid-rapidities and offering hadron, electron, photon and jet measurements and the muon spectrometer at forward rapidity providing muon detection.
The large rapidity domain covered by ALICE offers the possibility to study the QGP in different charged particle multiplicity density regimes.%

The experiment's requests to LHC, are to provide after \LSTwo, an integrated luminosity in \pb collisions of 10~\nbinv and higher, plus \pPb high-luminosity  and pp reference runs at 5.5~TeV. \\

The ALICE upgrade physics goals after \LSTwo can be split into three main objectives~\cite{Abelevetal:2014cna}.
The first one is the study of the heavy-quarks and in particular the thermalization of heavy quarks in the medium, and the parton mass and colour-charge dependence of in-medium energy loss.
These two topics are closely connected as the in-medium heavy-quark energy loss lowers the momenta of heavy quarks, they may thermalize in the system, and then participate in the collective dynamics of the medium. 
The simultaneous observation of the two phenomena opens the possibility for determination of the QGP transport coefficients.
This goal will be achieved by measuring the charm and beauty hadrons production and dynamics to the lowest \pt.\\
The second topic is the study of charmonia production in \pb collisions. 
Charmonium is the first hadron for which a clear mechanism of suppression in QGP was proposed, based on the colour-charge analogue of Debye screening.
Two alternative models were proposed, the first one is the statistical hadronization model in which the charm quarks produced in the initial hard collisions thermalize in the QGP and are distributed into hadrons at chemical freeze-out. 
The second one proposes the kinetic recombination of c and cbar quarks in the QGP as an alternative charmonium production mechanism. 
In this model a continuous dissociation and regeneration of charmonium takes place in the QGP over its entire lifetime. 
The first LHC data showed  that regeneration of \jpsi plays an important role at low \pt~\cite{Abelev:2013ila}.
The measurement of the production of different charmonium states in \pb collisions at the LHC will provide a definitive answer on the question of \jpsi production mechanism in the QGP.\\
Finally, the measurement of low-mass dilepton production gives an insight into the bulk properties and the space-time evolution of the QGP.
Also, The masses of the light-quark particles are connected with spontaneous breaking of chiral symmetry of QCD. 
The theory predicts that this fundamental symmetry is restored at high temperature, leading to substantial distortions of the spectral functions. 
Such modifications, in particular for the $\rho$ meson, should be observable in dilepton spectra. 
To achieve these goals, high statistics and high precision measurements are required.
Selective hardware triggers are not effective for low \pt and low S/B which impose to acquire a large statistics in minimum-bias mode by increasing the readout rate and reducing the data size. 
The ALICE goal is to be able to run at 50~kHz interaction rate in \pb collision, which is 100 times faster than \RunOne rate.
The planned upgrades will enhance the vertexing and tracking capabilities  while preserving the excellent particle identification capabilities of ALICE. 
 \section{\LSTwo UPGRADE OVERVIEW}
 
 To match these requirements, a new, high-resolution, low-material-thickness silicon inner tracker will improve the tracking resolution and will be extended to forward rapidities into the muon spectrometer acceptance.
In order to cope with the 50~kHz readout rate requirement, the readout of most of the present detectors will be upgraded.
The present V0/T0 detector system, presently used as MB triggers, will be replaced by a Fast Interaction Trigger detector (FIT), which will provide the Minimum Bias interaction trigger for the experiment.  
The FIT detector will consist of a new assembly of Cherenkov and scintillator detectors.
The Muon Trigger detector is at present providing the selection of high \pt single muon and di-muon events with a maximum trigger rate of 1~kHz. 
The upgrade trigger strategy does not foresee a muon trigger, all events will be read upon the interaction trigger and the  signals of the Muon Trigger detector are used offline for hadron rejection. 
Finally an integrated Online-Offline computing system, based on a new paradigm, is being developed to treat and record the acquired data.
 
 \subsection{Detector Upgrades}

A completely new all-pixel silicon  inner tracker detector will be installed covering the mid-rapidity detector as well as the forward muon spectrometer rapidity enabling vertexing at forward rapidity~\cite{Abelevetal:2014dna, CERN_LHCC_2015_001}.   
The chosen technology is the CMOS Monolithic Active Pixel Sensor (MAPS) in which the charge collection and the digitalization are done in the same silicon substrate.
It gives the possibility to thin down the sensors to 50~$\mu$m reducing the material budget. 
The pixel size can be reduced down to $28\times 28\;\mu{\rm m}^2$ to achieve an excellent spatial resolution of the order of 5~$\mu$m.
 And finally with a readout rate of 100~kHz in \pb collisions and 400~kHz in pp collisions, this technology complies with the readout requirements.
    \begin{figure}[h]
 \centerline{\includegraphics[width=0.5\textwidth]{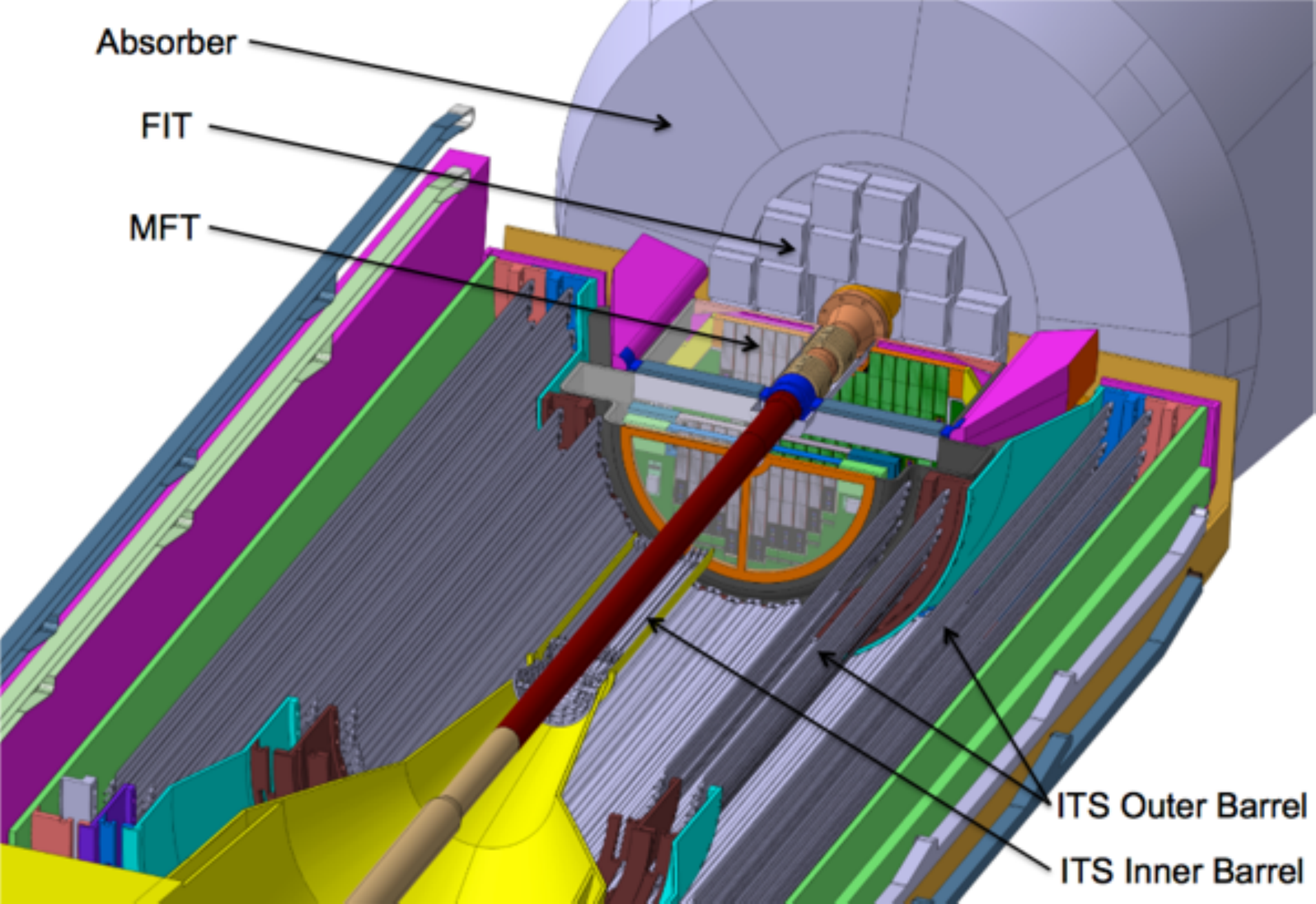}}
  \caption{Artistic view of the inner detectors (ITS, MFT and FIT) after \LSTwo.}
\end{figure}
\begin{figure}[h]
 \includegraphics[width=0.4\textwidth]{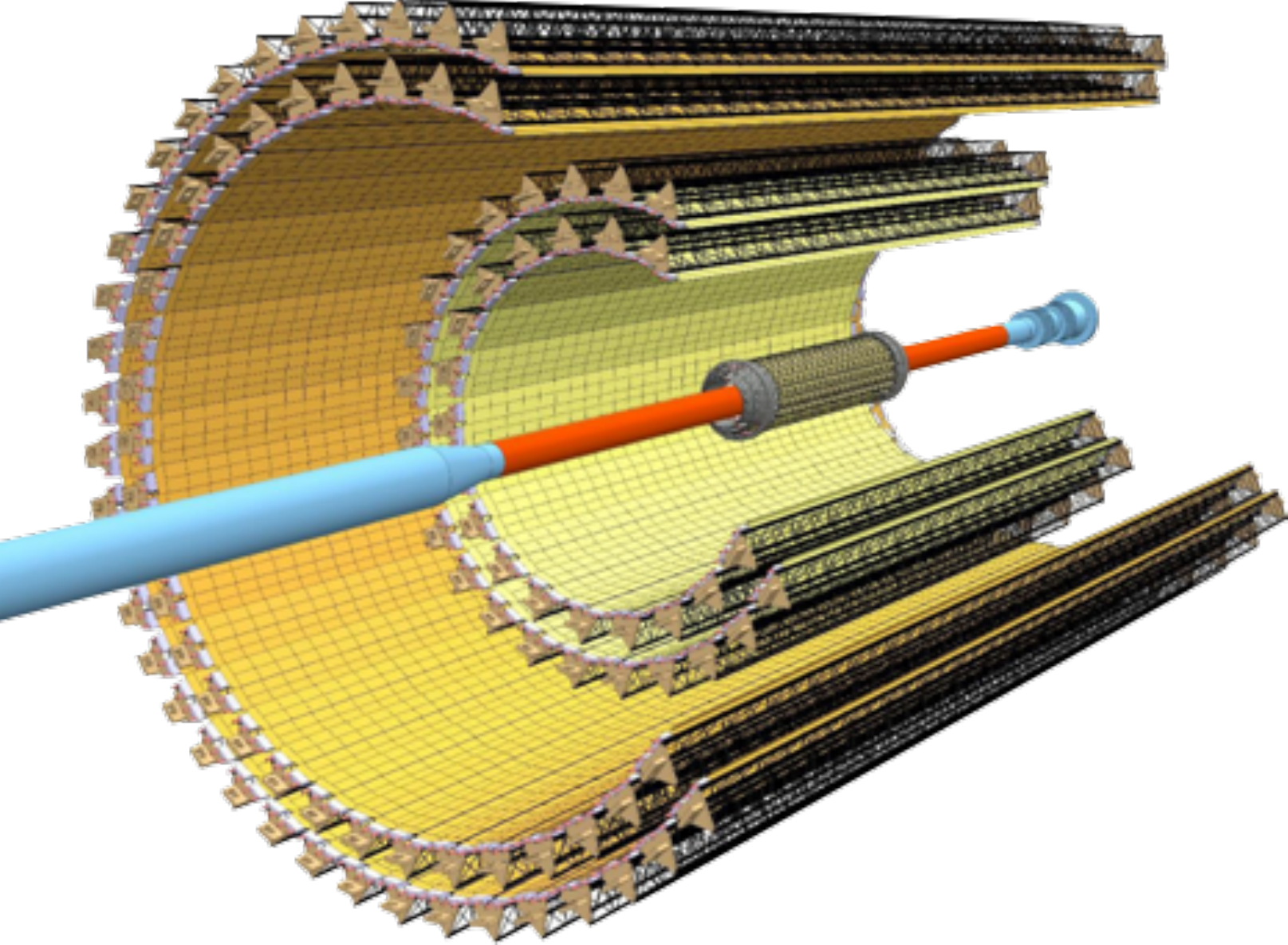} 
 \hspace{0.5cm}
 \includegraphics[width=0.35\textwidth]{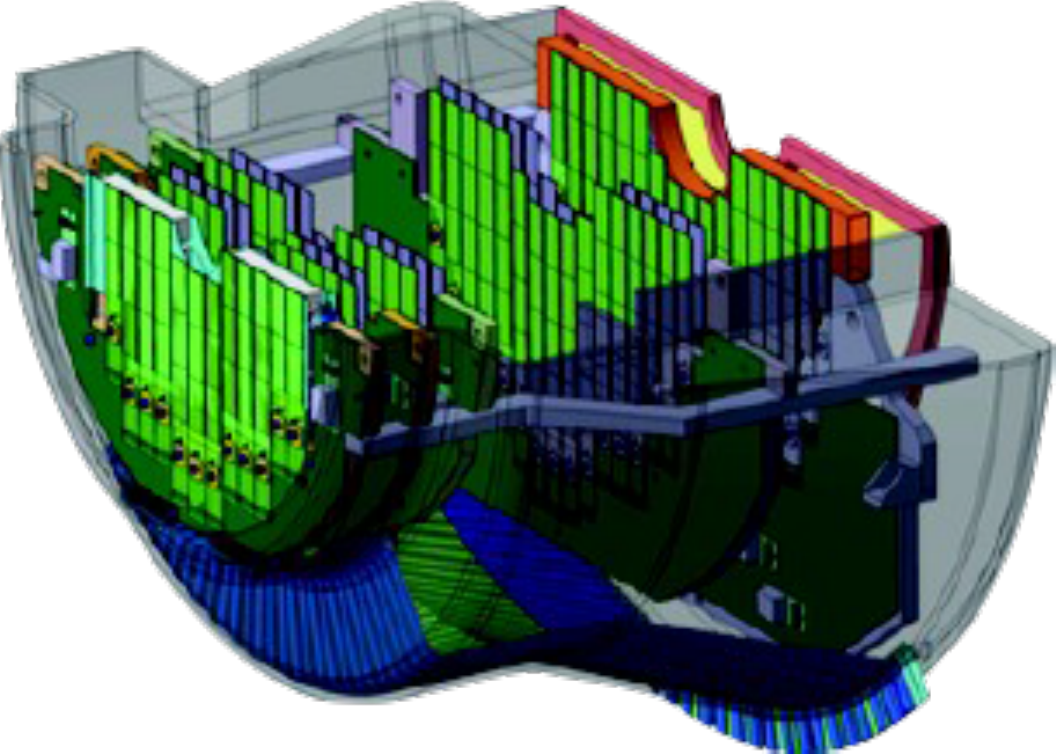} \\
  \caption{Detailed views of ITS (left) and MFT (right) detectors. }
\end{figure}

\begin{figure}[h]
 \includegraphics[height=0.35\textwidth]{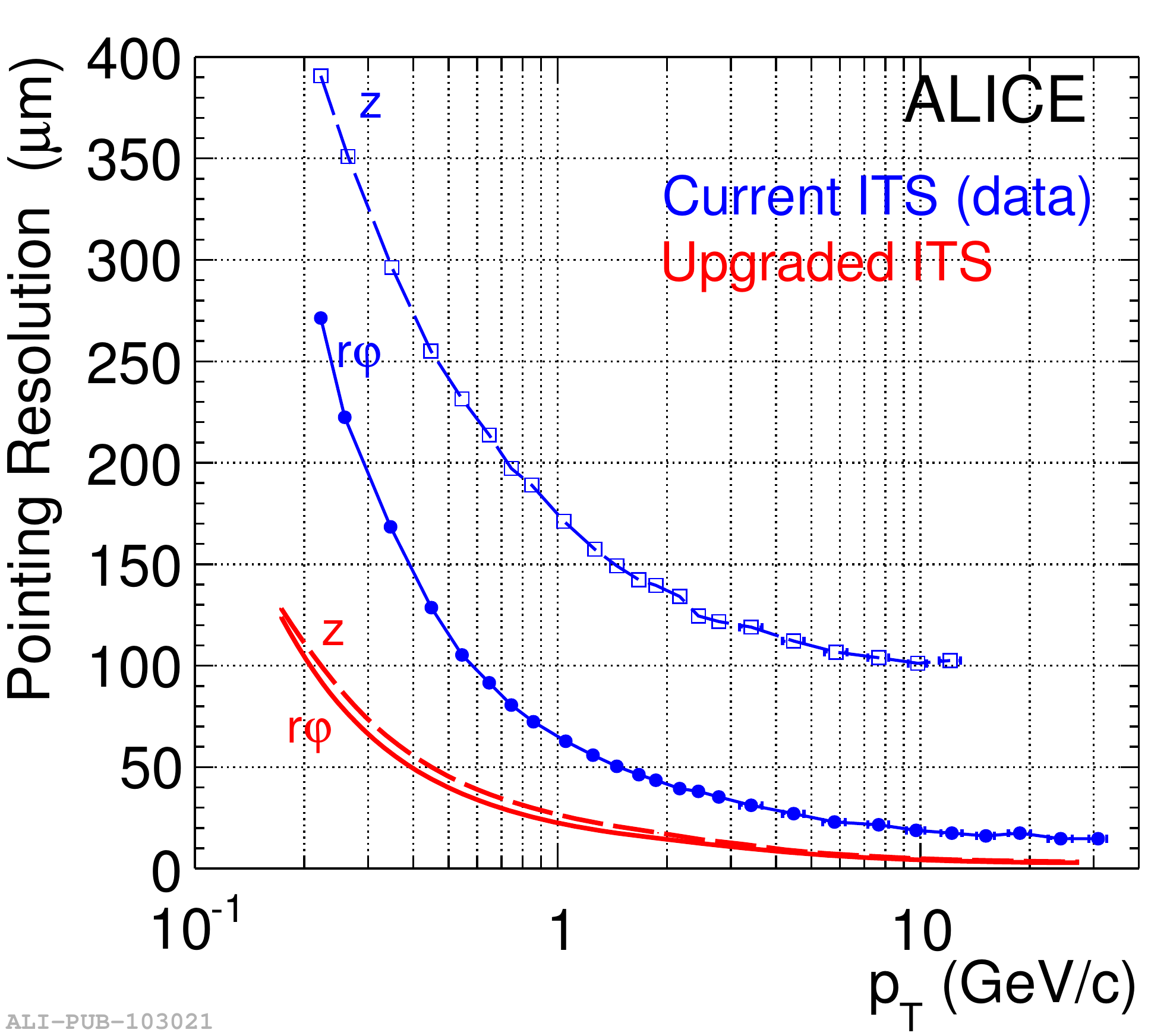} 
 \hspace{0.5cm}
 \includegraphics[height=0.35\textwidth]{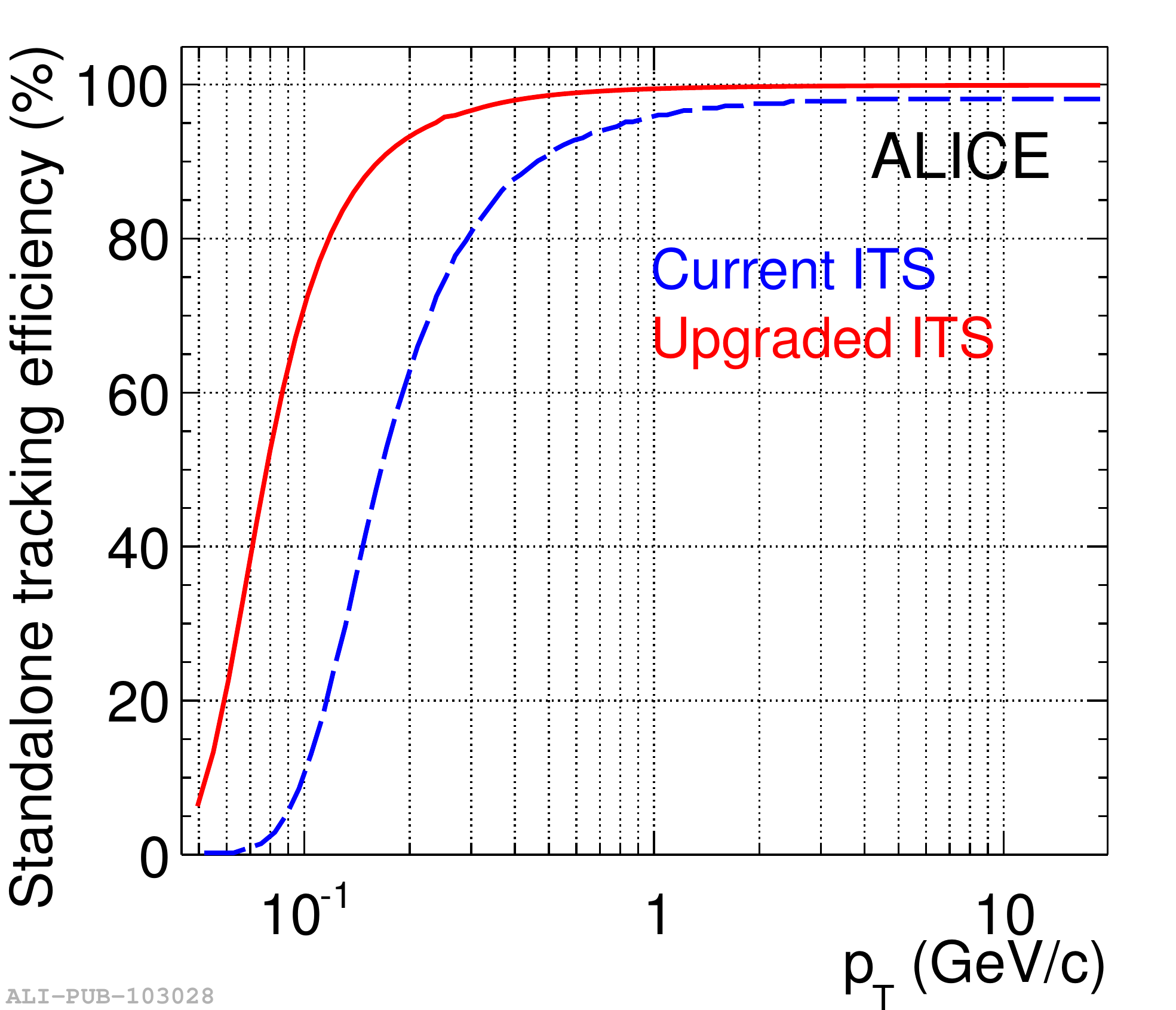} \\
  \caption{Left: Pointing resolution of the new ITS (red lines) compared to the present apparatus (blue lines and markers) in both longitudinal direction ($z$) and transverse plan (r$\varphi$) versus \pt.
  Right: Track reconstruction efficiency  of new ITS (red line) compared to the present apparatus (blue dashed line) versus \pt~\cite{Abelevetal:2014dna}. }
  \label{fig:ITSPerf}
\end{figure}%
\begin{figure}[h]
 \includegraphics[height=0.35\textwidth]{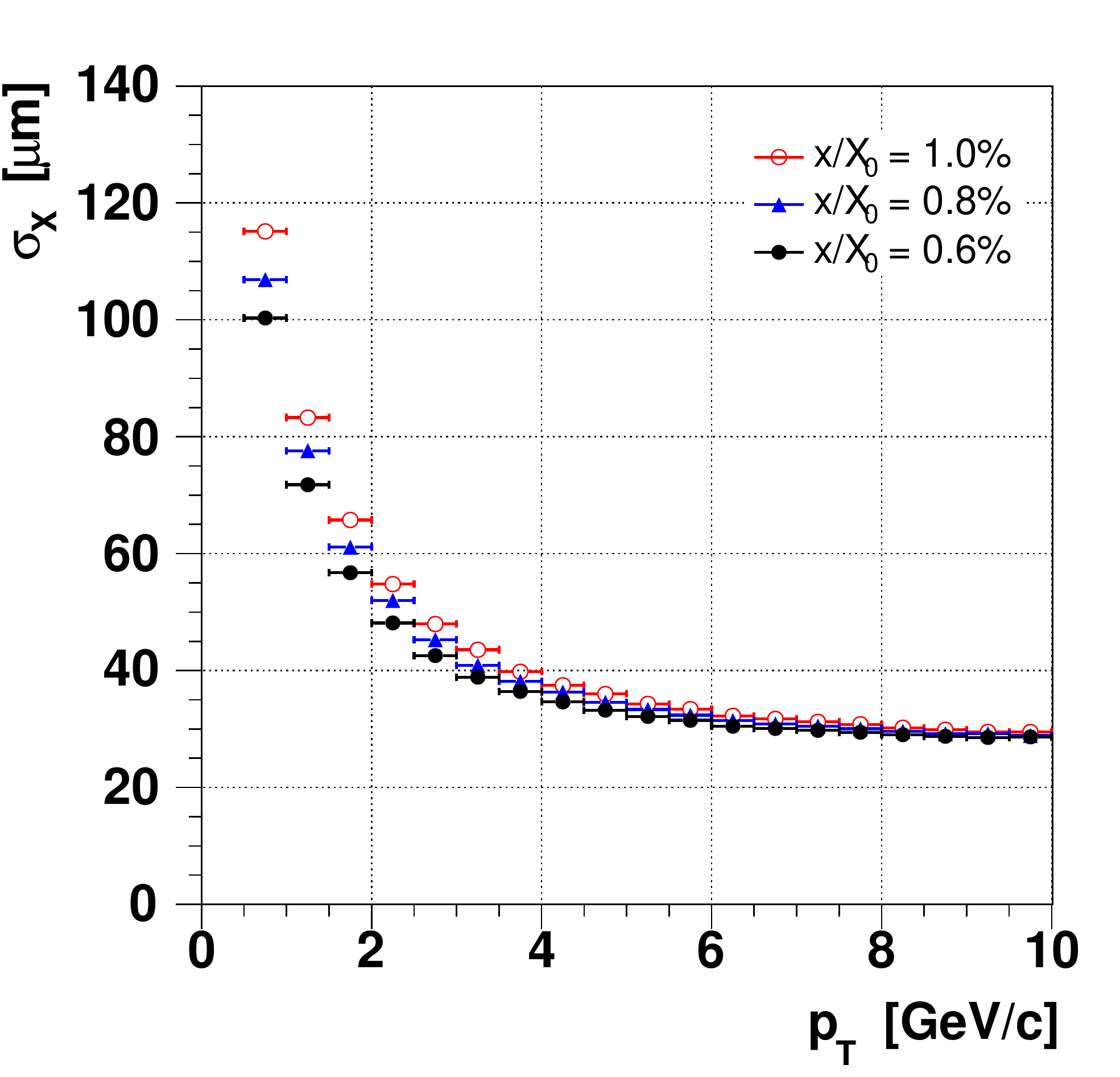} \\
 \includegraphics[height=0.35\textwidth]{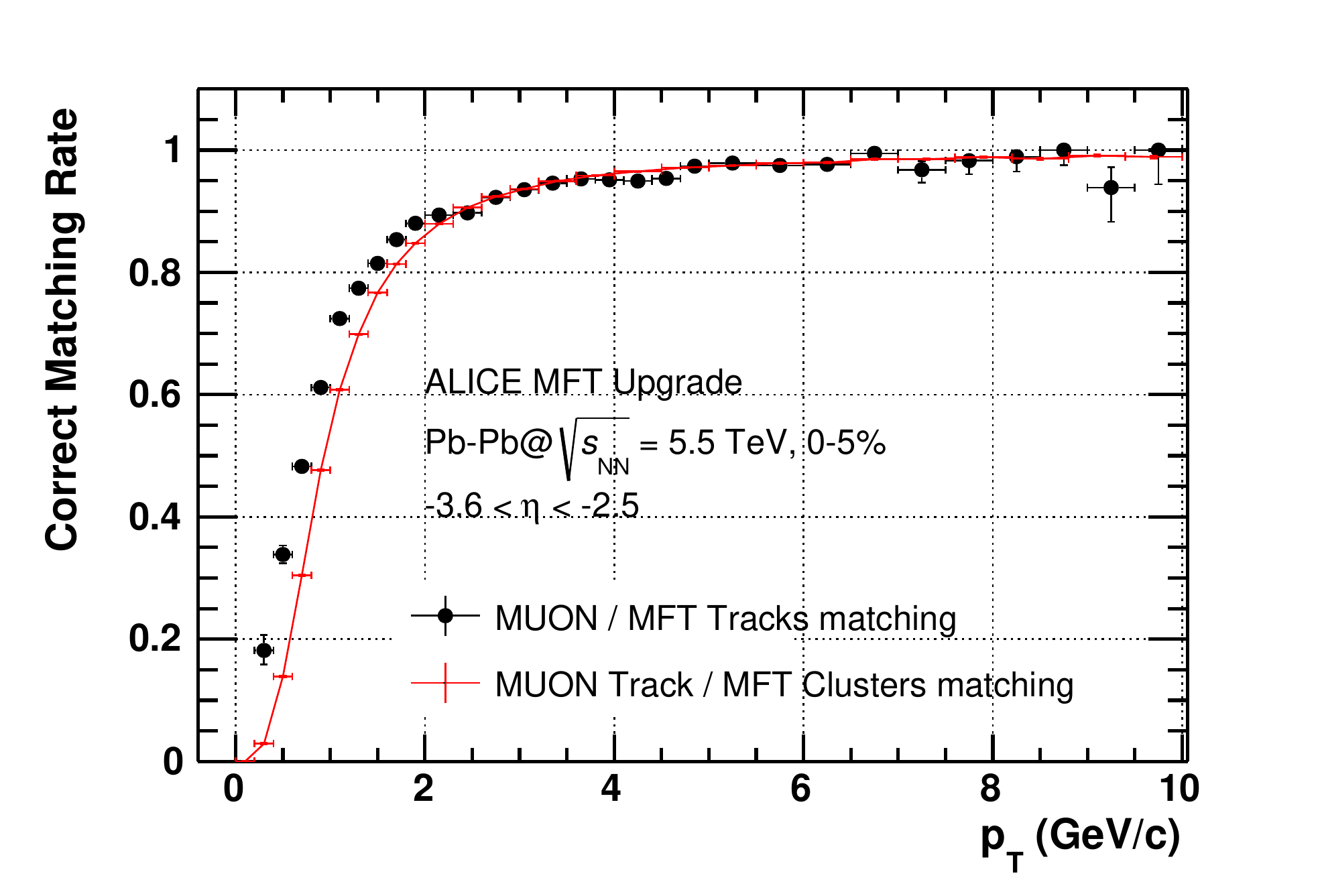} 
  \label{fig:MFTPerf}
  \caption{Left: Pointing resolution of the MFT in the transverse plan versus \pt, different marker/color show the sensitivity of the pointing resolution to the material budget per layer. 
  Right: Correct matching rate between MFT and MUON tracks in central \pb collisions versus \pt. Two different algorithms were tested providing similar performances~\cite{CERN_LHCC_2015_001}.}
\end{figure}
The ITS at mid-rapidity~\cite{Abelevetal:2014dna} will consist in 7 layers of CMOS MAPS, 3 inner layers, 2 middle layers and 2 outer layers.
The rapidity coverage  ($|\eta|<1.5$) will be extended by more than one unit of rapidity compared to the present ITS ($|\eta|<0.9$).
A new design of the beam pipe with a smaller radius ($r=18.6$~mm) allows the first detection layer to be placed closer to the interaction point (IP) at a radius of 23~mm compared to 39~mm with the present apparatus.
The CMOS MAPS technology allows  reduction of the detector material budget  which will be as low as 0.3\% of $X_0$ for the inner layers and 0.8\% of $X_0$ the other layers, making the the ALICE inner tracker the thinnest at LHC.
As seen on the left panel of Figure~\ref{fig:ITSPerf}, the impact parameter resolution will be improved by at least a factor 3 in the transverse plane and a factor of 5 along the beam axis for \pt $>$ 0.5~GeV/$c$.
The right panel of Figure~\ref{fig:ITSPerf} presenting the track reconstruction efficiency versus \pt, shows that lower \pt will be reached with an excellent tracking efficiency of 90\% down to a \pt of about 0.1--0.2~GeV/$c$.
In the forward rapidity domain, ALICE detects muon thanks to the muon spectrometer.
A thick absorber of about 60 radiation lengths offers muon identification by absorbing hadrons. 
The drawback is that the
multiple scattering of muons in the absorber introduces a large uncertainty
in extrapolation of the tracks to the interaction point and then does not allow secondary vertex measurement which is crucial for charm/beauty separation in \jpsi or single muon analyses. 
The MFT~\cite{CERN_LHCC_2015_001} which is the forward part of the inner silicon tracker, and standing for Muon Forward Tracker, will increase the pointing accuracy of muon tracks by matching the MUON spectrometer tracks with the one measured in the MFT before they go through the absorber.
The MFT consists of 5 disks placed perpendicular to the beam axis with each disk composed of 2 detection planes of CMOS MAPS covering a pseudorapidity of $-3.6<\eta<-2.3$.
A material budget of 0.6\% of $X_0$ per disk will be achieved.
The left panel of Figure~\ref{fig:MFTPerf} shows the impact parameter resolution in the transverse plane versus \pt, shows that values better than 100~$\mu$m will be achieved for $\pt > 1$~GeV/$c$ which is the lowest single muon \pt used to study quarkonia and heavy-flavour in the muon spectrometer.
Right  panel of Figure~\ref{fig:MFTPerf} shows that a correct MFT/MUON matching rate greater than 70\% for $\pt > 1$~GeV/$c$ in central \pb collisions is met.
Those performances allow  charm/beauty separation in single muon and \jpsi analyses via the measurement of secondary vertex in the case of \jpsi from B-hadron decays or via a closest approach measurement for heavy-flavour studies in their single muon decay channel.\\

The present Time Projection Chamber (TPC) readout based on multi-wire proportional counter chambers is limited to a rate of about 3~kHz by the ion back flow.
New readout chambers based on the GEM technology using a quadrupole GEM configuration will be used, reducing this effect and will allow the TPC  to be operated  with a continuous readout up to 50~kHz in \pb collisions~\cite{CERN-LHCC-2013-020, CERN-LHCC-2015-002}.
The momentum resolution for tracks combing the TPC and ITS information is preserved compared to the present apparatus. 
The same applies to the excellent PID efficiency via d$E$/d$x$ measurement which will be essential to enhance the significance of rare observables measured after \LSTwo that suffer from a low signal-to-background ratio.

 \subsection{Readout/Trigger Upgrade}

\begin{figure}[h]
 \includegraphics[width=0.7\textwidth]{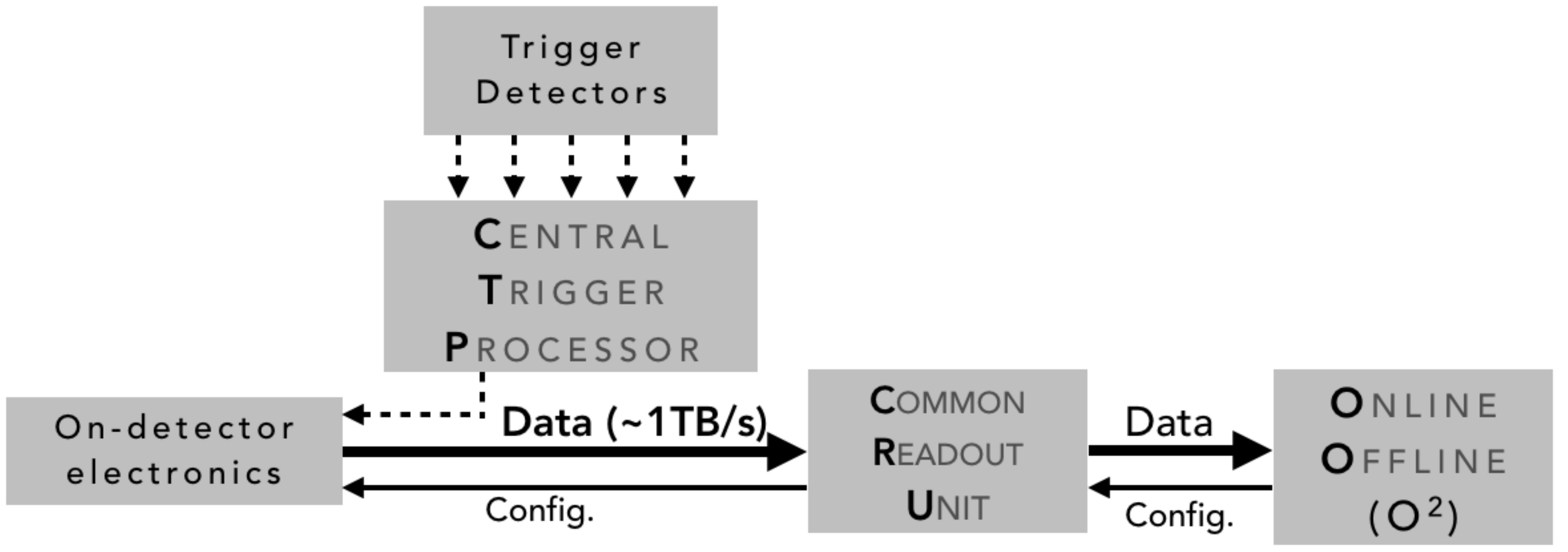}
  \label{fig:Readout}
  \caption{Schematic description of the readout and trigger system.}
\end{figure}

The goal of the readout system is to be able to record all \pb collisions at 50~kHz and 200~kHz in pp and p--Pb~\cite{Antonioli:1603472}.
Figure~\ref{fig:Readout} shows a schematic of the readout architecture.
Detectors will be readout upon a minimum-bias trigger (provided by the FIT detector) or be readout continuously.
The central trigger processor (CTP) will be upgraded to accommodate the higher interaction rate, providing trigger and timing distribution (TTS) to the upgraded detectors and backwards compatibility to detectors not upgrading their TTS interface.
The readout of the TPC detector as well as the muon chambers will be performed by a dedicated ASIC called SAMPA. 
The SAMPA chip will contain 32 channels and is based on the developments of the actual readout chip of the TPC, ALTRO and S-ALTRO. 
A new Common Readout Unit (CRU) is being developed and will be used by most of the detectors. 
The CRU  provides interface between the detector FEE and Online computing.
Data transmission between detector Front End Electronics and the CRUs will use GBT links.
A total data flow of about 1~TB/s in \pb running mode has to be readout of the detectors which represents a data throughput approximately 100 times larger than the one achieved during Runs 1 and 2. 
This high data throughput requires a change in the paradigm of the Online/Offline system.

 \subsection{New Online-Offline Computing System}
\begin{figure}[h]
 \includegraphics[width=0.7\textwidth]{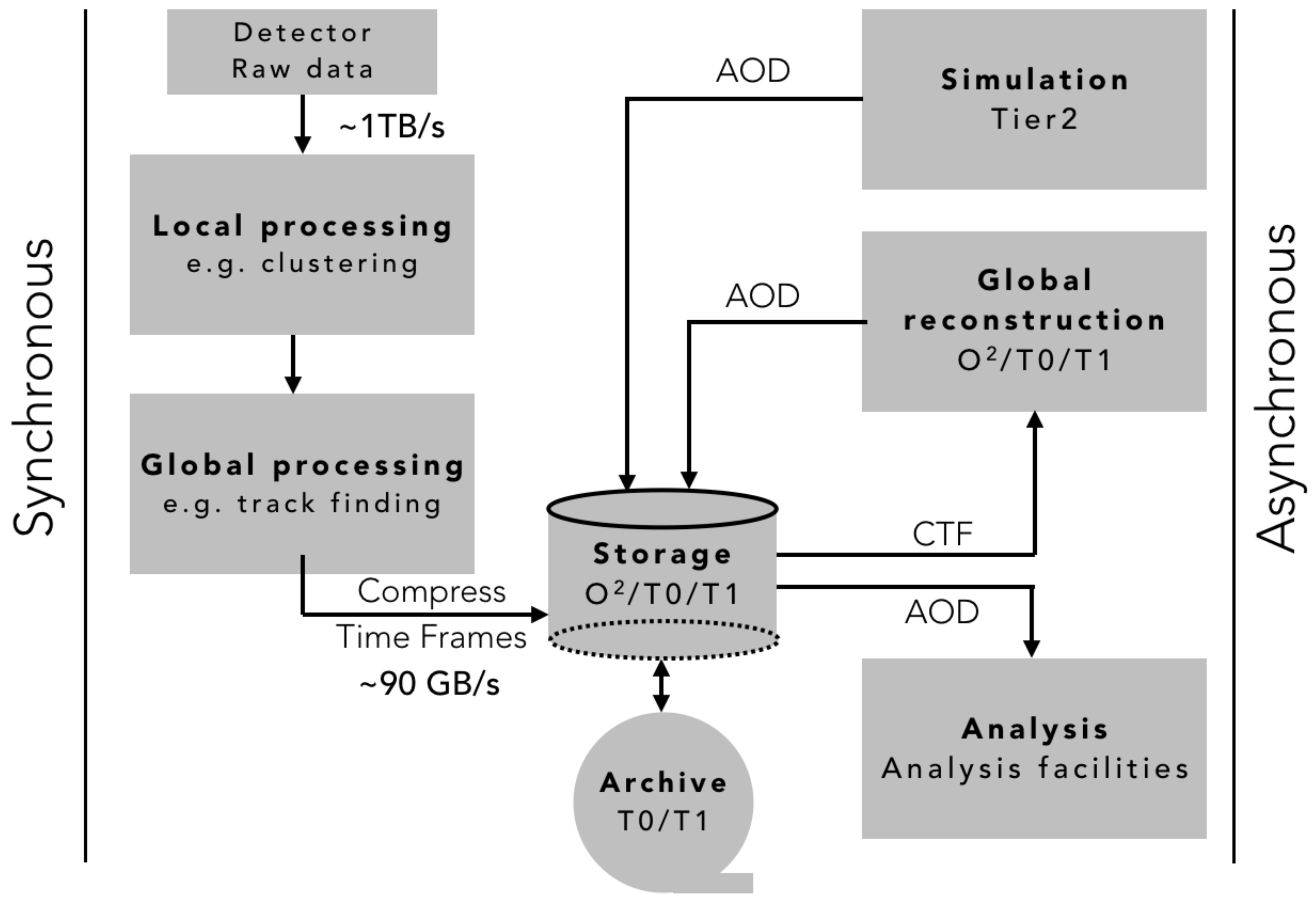}
  \label{fig:O2}
  \caption{Schematic description of the new Online and Offline computing system.}
\end{figure}

A new facility at Point 2 named O$^2$ standing for Online-Offline computing system is being developed~\cite{Buncic:2011297}.
It will consist in 100000 CPU cores coupled to 60~PB of storage capacity.
The goal of this system is to allow a maximum data reduction as early as possible in the data flow.
This is achieved by a complete change in the data acquisition and reconstruction paradigm.
Two stages of reconstruction will be performed, one synchronously with the raw data acquisition and the second asynchronously (see Figure~\ref{fig:O2}).
The synchronous reconstruction consists in a first step of local processing of the standalone detector data.
One of the main goals of the local processing is the data reduction, in particular for the TPC.
During global processing, the information of the full ALICE detector is available as input. 
Global processing includes detector reconstruction, calibration and quality control.
The data being stored as Compressed Time Frame format has a flow reduced to about 90 GB/s.

The final global data reconstruction is performed asynchronously, using the O$^2$ facility if CPU power is available or via the same scheme as during Runs 1 and 2 which is using the Tier-0 and 1 of the LHC grid.
 The final user analyses will be performed in a small number (2-3) of analysis facilities.
 The chosen sites will have very good network connectivity and a large, efficient local disk buffer dedicated to the storage of an entire set of active ALICE AODs.
The Tier-2 centres of the LHC grid will be dedicated to the physics simulation production.

 \section{PHYSICS PERFORMANCES -- SELECTED HIGHLIGHTS}

The ALICE physics programme for Runs 3 and 4 is large and diverse. 
Its complete description can be found in the ALICE upgrade's Letter of Intent and its addendum~\cite{ALICE_Upgrade_LoI,LoI_Addendum}.
In this presentation, we focused on the physics performance description related to the open heavy-flavour and quarkonia studies.
High-precision measurements of charm and beauty production in heavy-ion collisions at the LHC is one of the principal physics motivations for the upgrade of the ALICE detector.
Heavy quarks play a special role in heavy-ion physics as they constitute an identified probe from their production  in the early collision stage via hard processes to their observation.
They are therefore sensitive to the full medium evolution which enables a unique access to their interactions in the QGP, allowing  insights to gained  into partonic energy loss, hadronization and thermalization processes and recombination mechanisms.
The ALICE upgrades will tremendously improve the accuracy of existing observables and make available new ones shedding new light on the QGP properties.\\

\begin{figure}[h]
\begin{tabular}{cc}
 \includegraphics[height=0.35\textwidth]{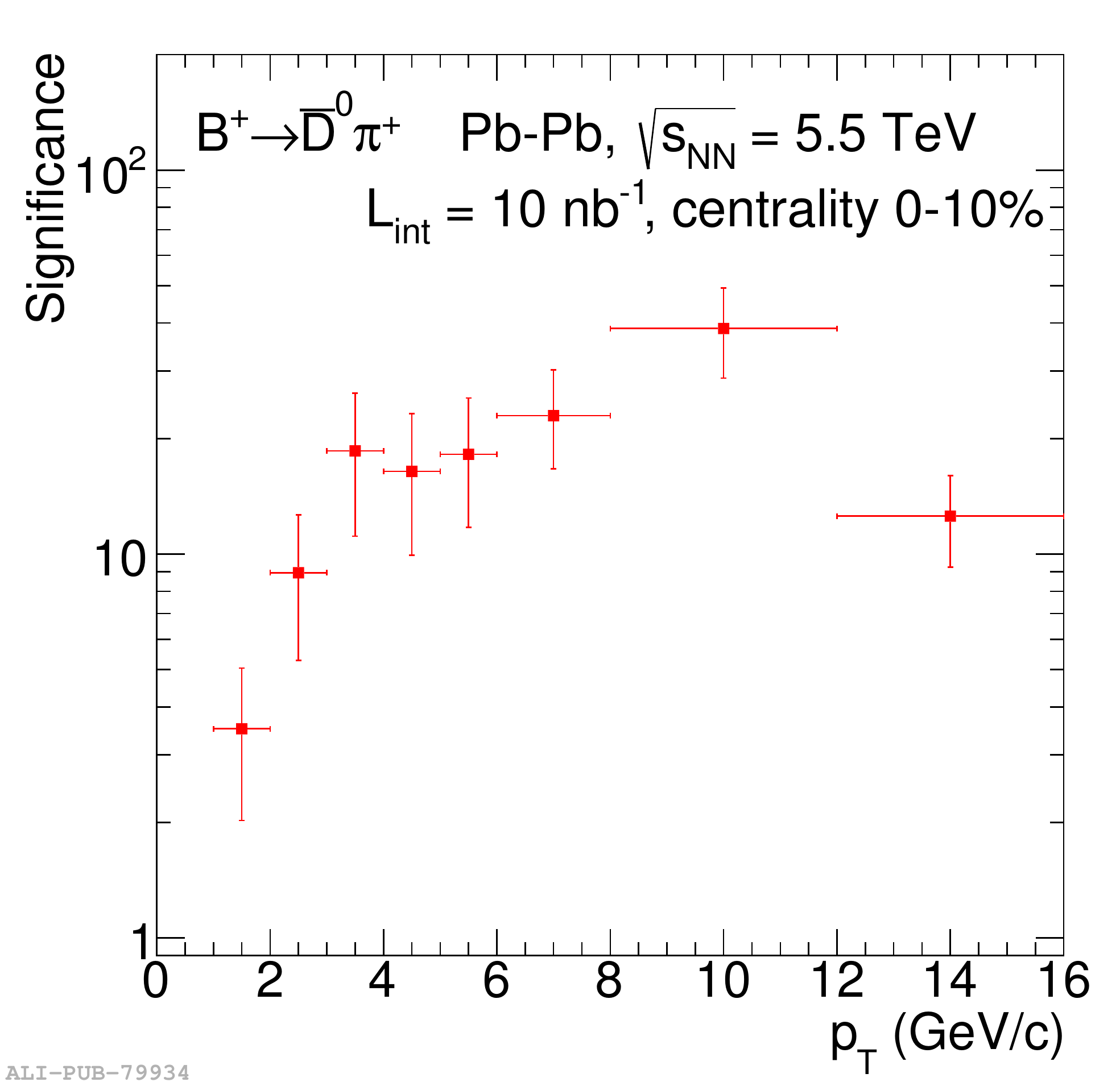} &
 \includegraphics[width=0.47\textwidth]{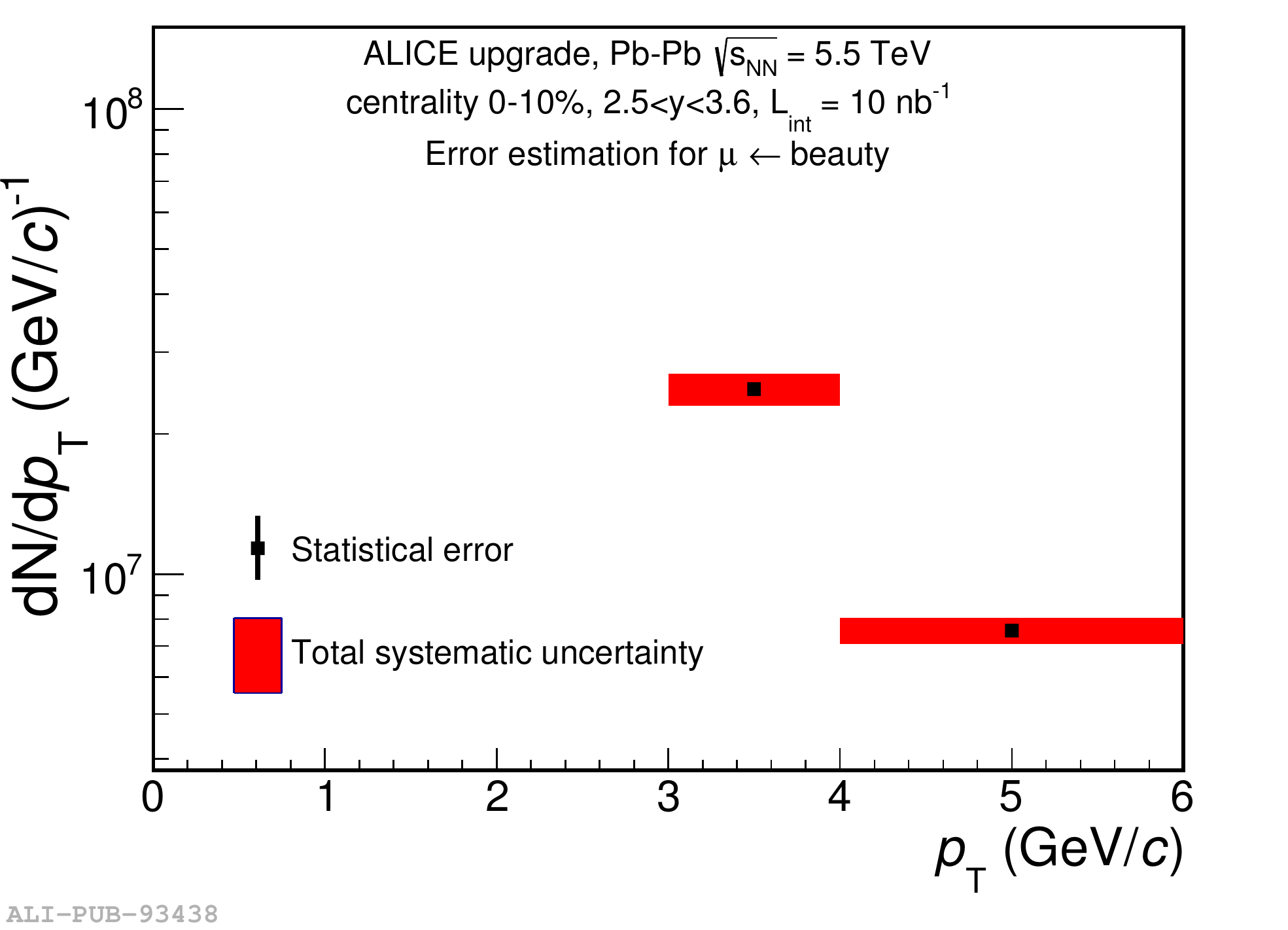} \\
  \end{tabular}
  \label{fig:Perf:Beauty1}
  \caption{For the most central \pb collisions centralities, significance of B$^+$ identification via its $\bar{\rm D}^0\pi^+$ decay channel at mid-rapidity versus \pt of the measured decay product (D$^0$ or muon)~\cite{Abelevetal:2014dna} (left) and estimated statistical and systematic errors on the beauty yield extraction at forward rapidity in single muon analysis for the lowest 2 \pt bins~\cite{CERN_LHCC_2015_001} (right).  }
\end{figure}

\begin{figure}[h]
\begin{tabular}{cc}
 \includegraphics[height=0.35\textwidth]{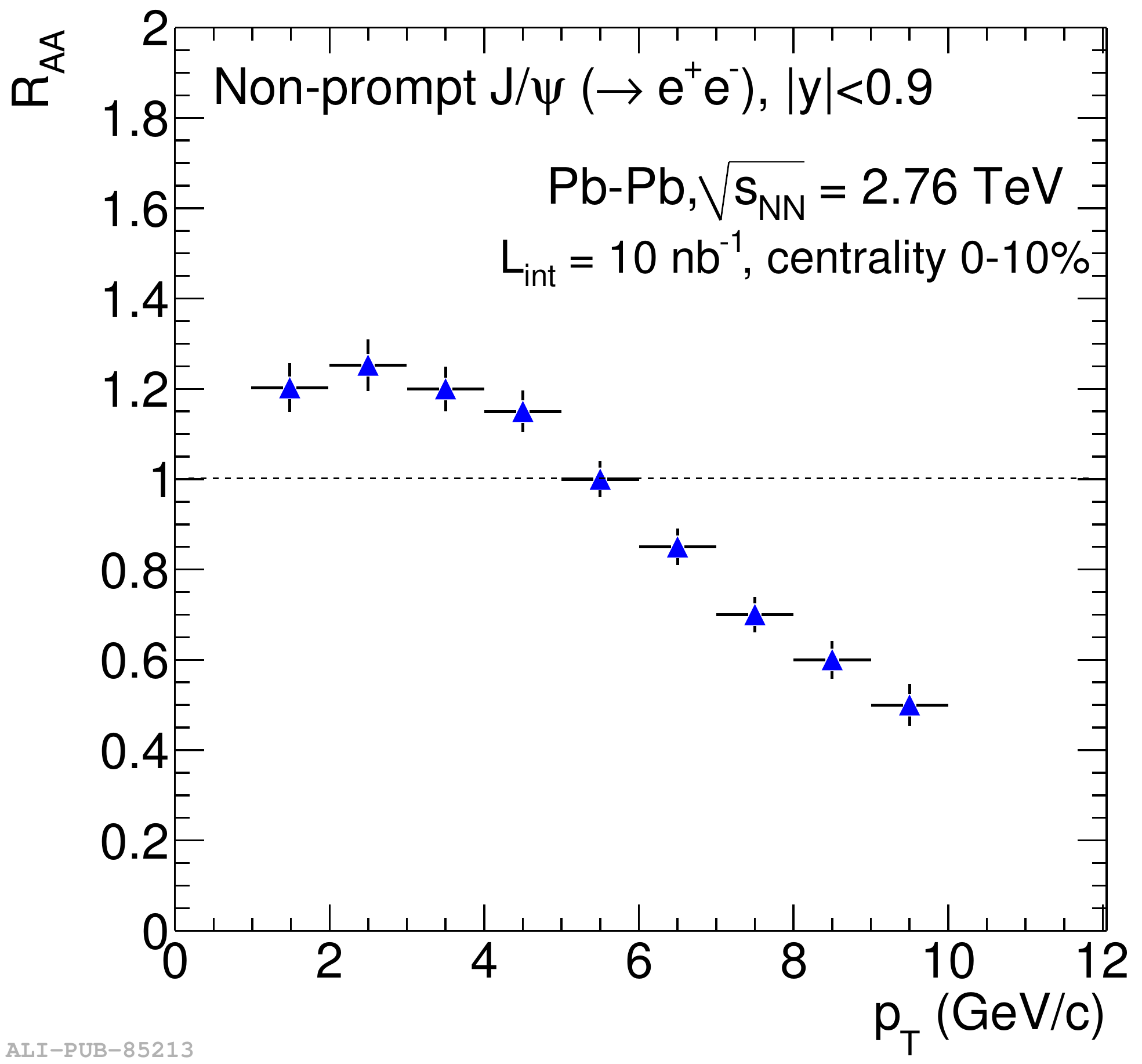} &
 \includegraphics[height=0.37\textwidth]{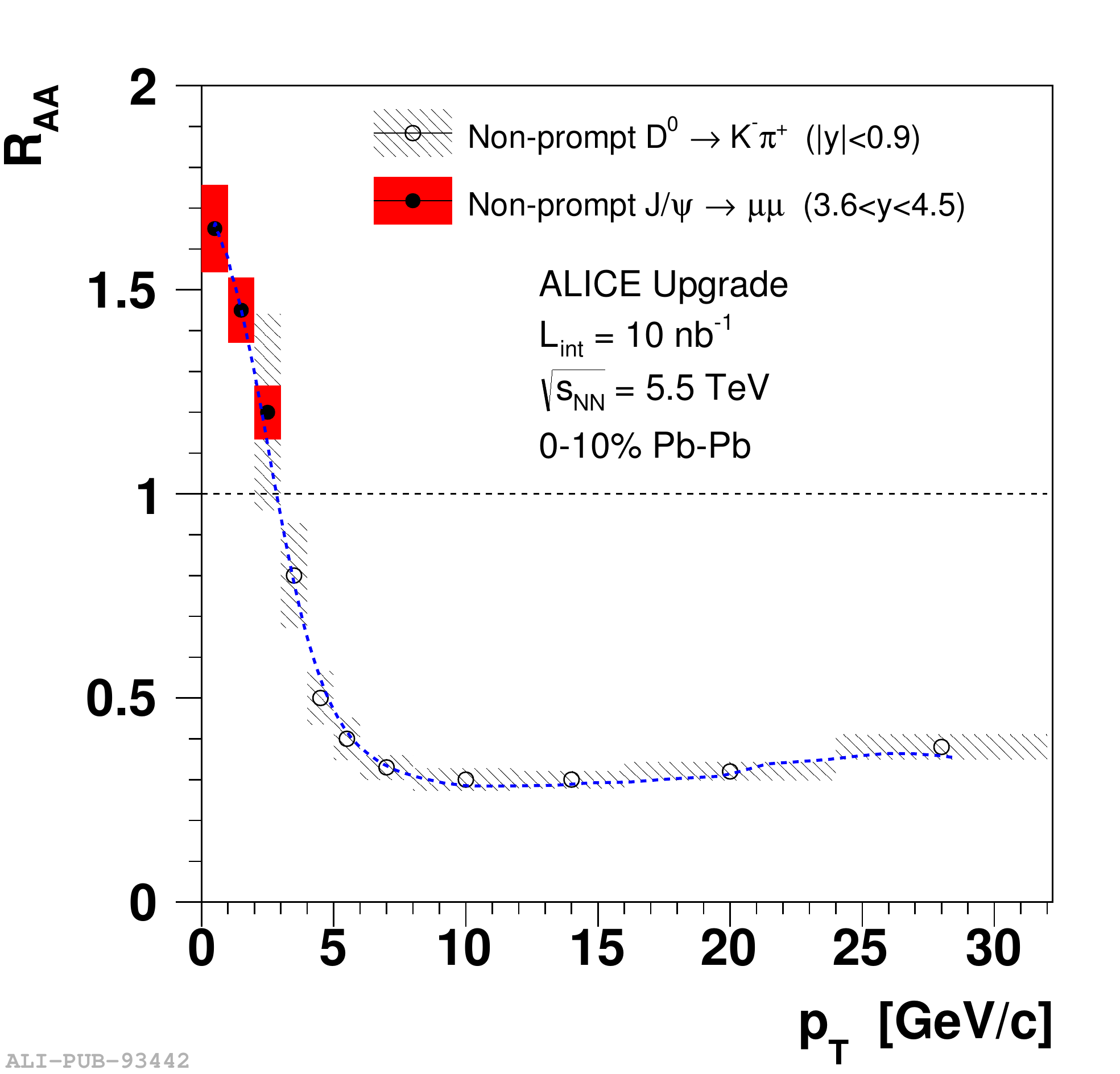} \\ 
 \end{tabular}
  \label{fig:Perf:Beauty2}
  \caption{Estimated errors on the beauty $R_{\rm AA}$ extraction using non-prompt \jpsi measured in its di-electron decay channel at mid-rapidity~\cite{Abelevetal:2014dna} (left) and in its di-muon decay channel at forward rapidity~\cite{CERN_LHCC_2015_001} (right) versus \pt of the measured \jpsi. }
\end{figure}

The nuclear modification factor, $R_{\rm AA}$, defined as the ratio of yield measured in \pb collisions by the one measured in pp scaled by the number of binary nucleon collisions is the tool to study production modification in \pb collisions compared to the pp reference.
A dependence of in-medium energy loss on the parton species would result in an ordering of the $R_{\rm AA}$ of beauty, charm and lighter quarks.
For the first time, combining CMS and ALICE results, a quark mass ordering of the in-medium parton energy loss is suggested by the data~\cite{Adam:2015rba}.
However, this observation is still limited to a high \pt domain and the limited accuracy prevents making any final conclusion.
After LS2, in addition to the charm production study, ALICE will be able to access beauty production with a good precision down to zero \pt  of the B-hadron.
In particular for the first time ALICE will study open beauty production at forward rapidity.
The beauty measurement will be possible both  at mid- and forward rapidity using non-prompt \jpsi from B-hadron decay via dielectron and dimuon channels as well as semi-leptonic and hadronic decays of B-hadrons (see Figure~\ref{fig:Perf:Beauty1} and \ref{fig:Perf:Beauty2}). 

\begin{figure}[h]
\begin{tabular}{cc}
 \includegraphics[height=0.35\textwidth]{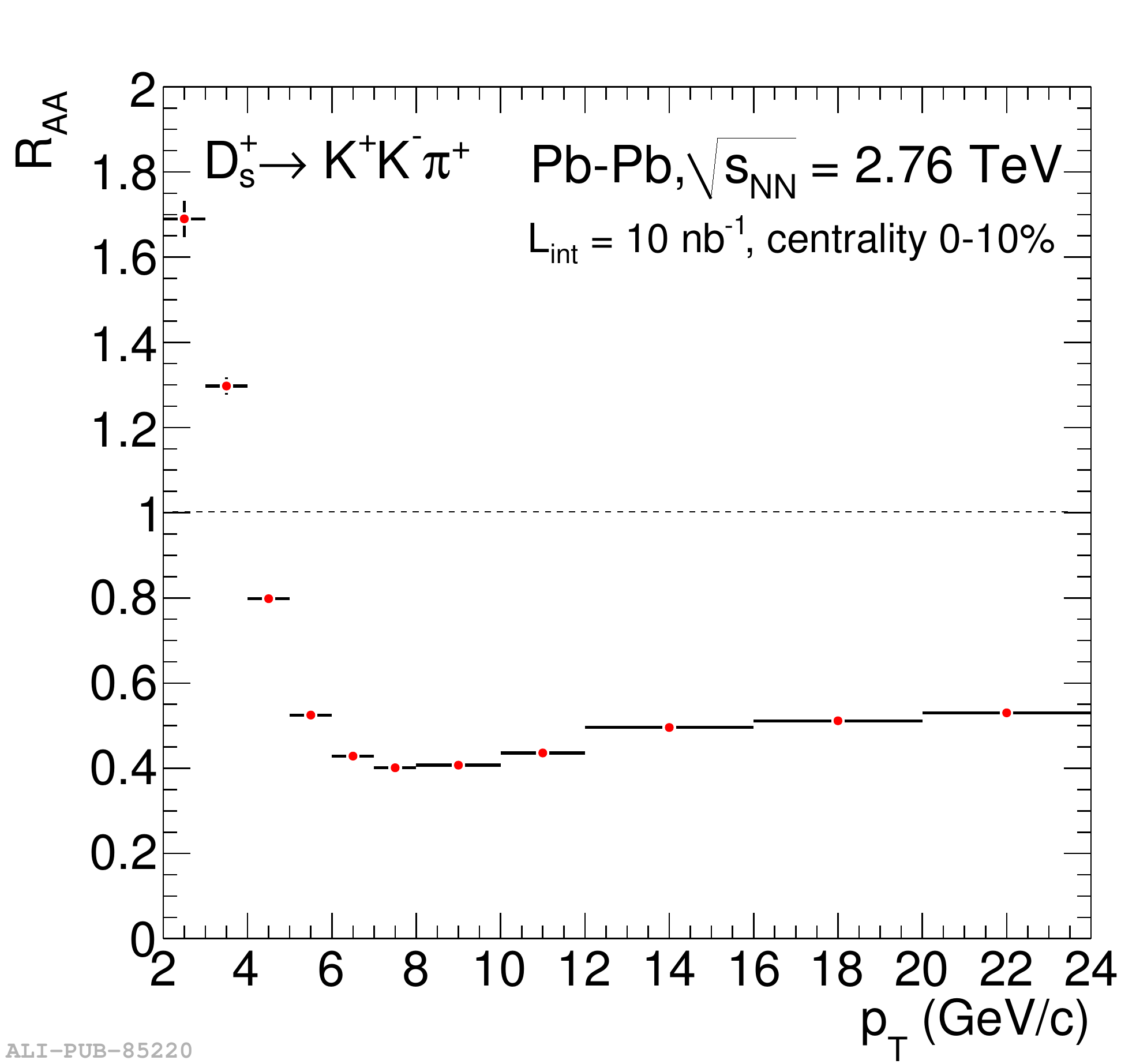} &
 \includegraphics[height=0.35\textwidth]{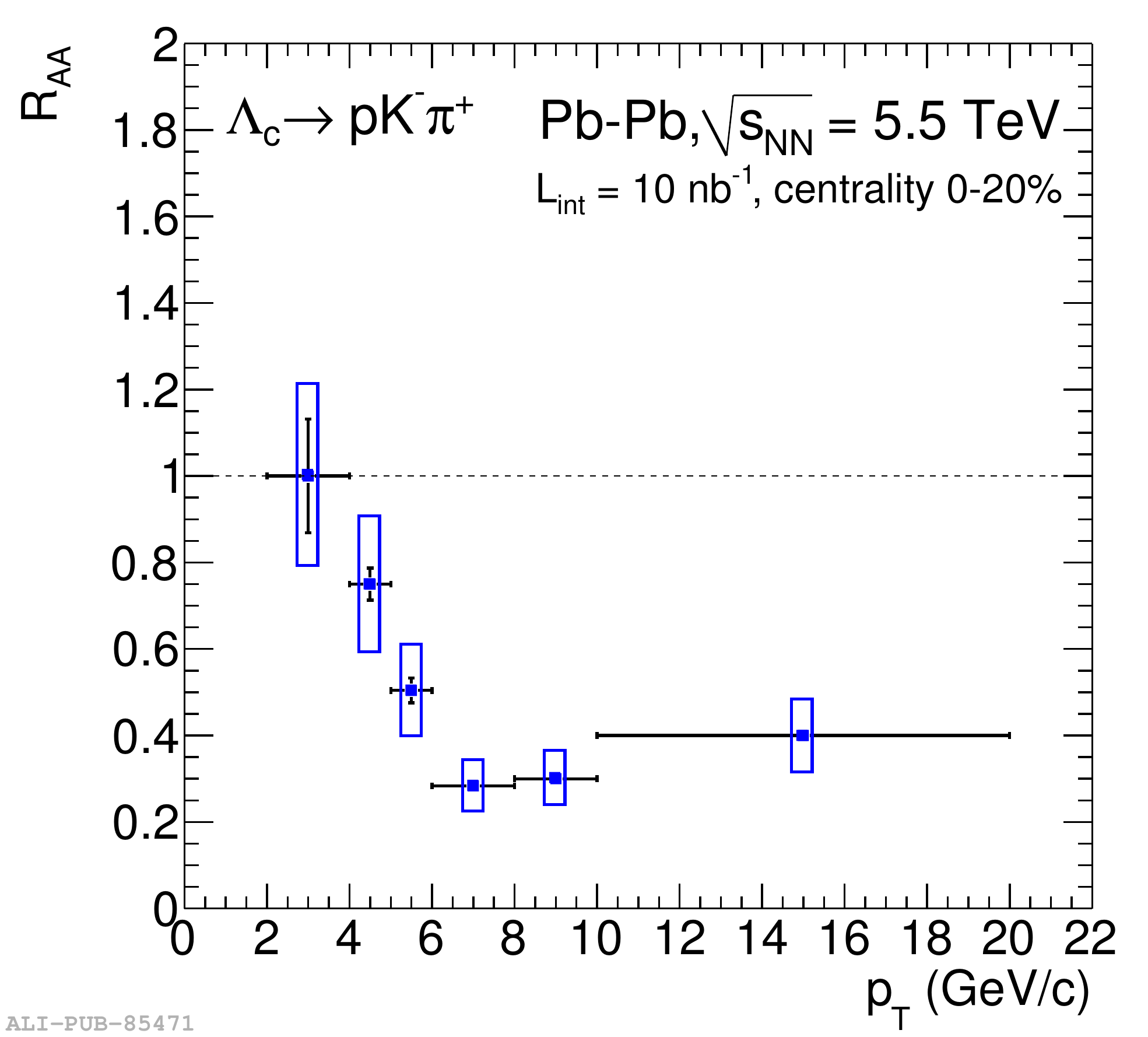} \\ 
 \end{tabular}
  \label{fig:Perf:Charm}
  \caption{Estimated errors on  $R_{\rm AA}$ extraction of ${\rm D}_s^+$ (left) and $\Lambda_c$ (right) at mid-rapidity in the most central \pb collisions~\cite{Abelevetal:2014dna}. }
\end{figure}

\begin{figure}[h]
\begin{tabular}{cc}
 \includegraphics[height=0.35\textwidth]{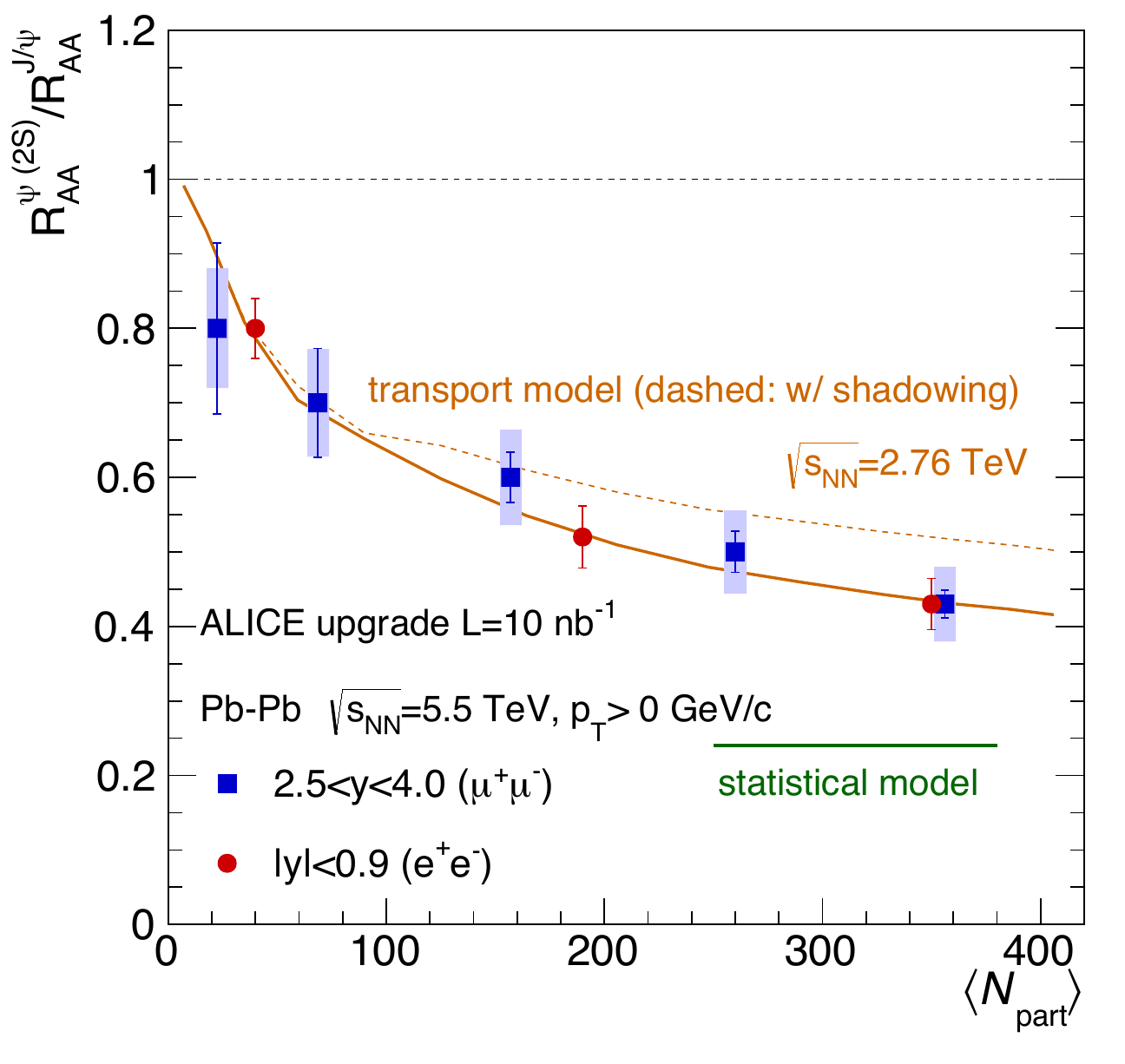} &
 \includegraphics[height=0.35\textwidth]{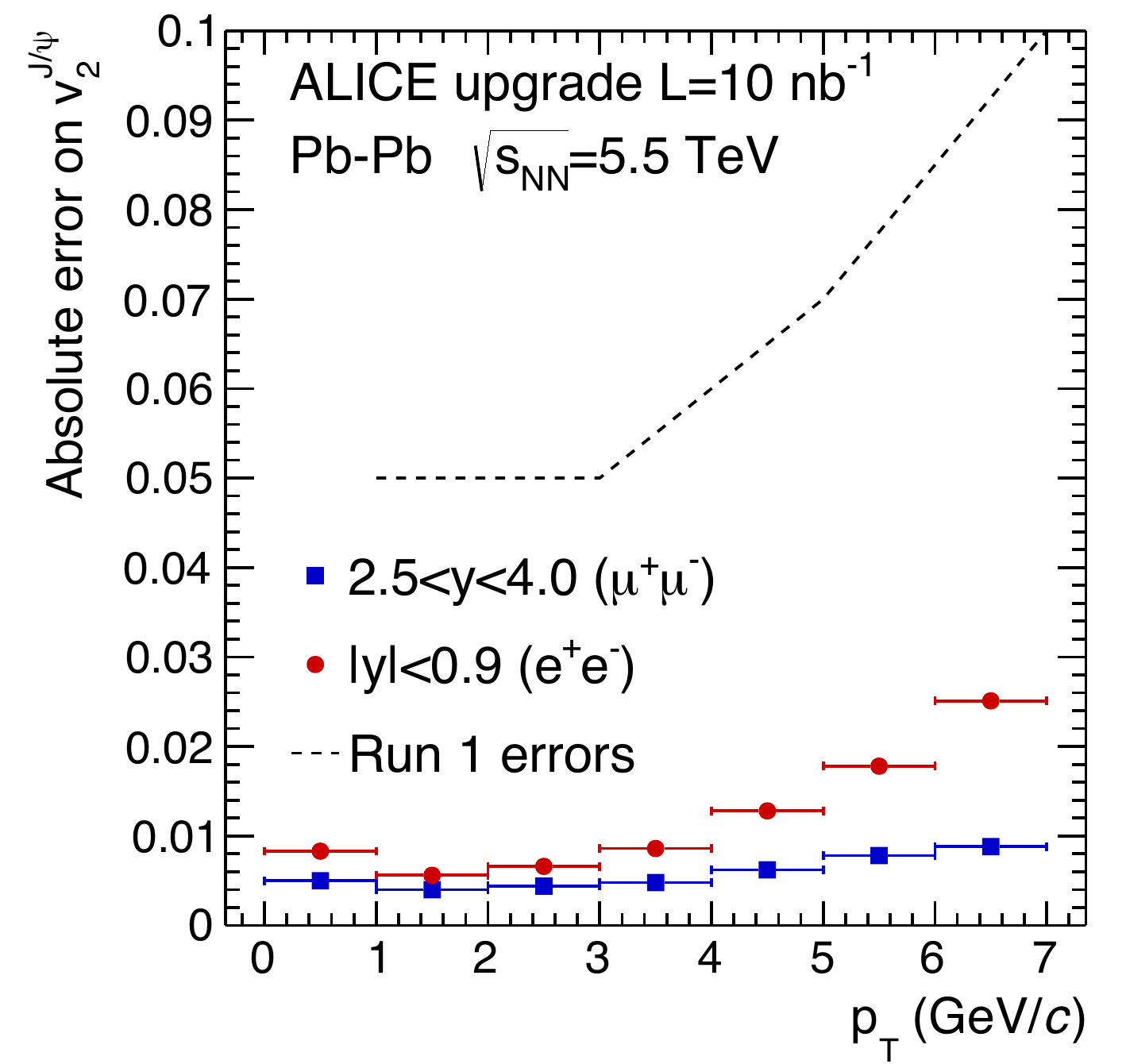} \\ 
 \end{tabular}
  \label{fig:Perf:jpsi}
  \caption{Left: Estimated errors on  $\psi$(2S)/ \jpsi ratio at both mid- and forward rapidity versus the number of participant nucleons. 
  Right: Estimated errors on \jpsi $v_2$ extraction at mid- and forward rapidities compared to actual precision. }
\end{figure}

Already available with the present detectors~\cite{ALICE:2012ab,Adam:2015jda}, charm production study in \pb collision will see a tremendous improvement of its precision and accuracy, together with an extension to zero \pt of the measurements for most of the D mesons.
The left panel of Figure~\ref{fig:Perf:Charm} shows as example the estimated error and \pt coverage for the  ${\rm D}_s^+$ nuclear modification factor measurement.
$\Lambda_c$ production measurement will be accessible for the first time in \pb collisions (see right panel of Figure~\ref{fig:Perf:Charm}).\\
The heavy-flavour dynamics will also be studied.
The standard way to characterize anisotropic flow uses a Fourier expansion of the azimuthal distribution of the outgoing particles.
The second coefficient of the expansion, $v_2$, is called elliptic flow.
The important question of thermalization of heavy quarks appears to be partly answered for charm where a positive elliptic flow is observed, which indicates that charm quarks take part in the collective expansion of the medium~\cite{Abelev:2013lca}. 
 But for the beauty sector, thermalization remains an open issue.
 \begin{figure}[h]
 \includegraphics[height=0.4\textwidth]{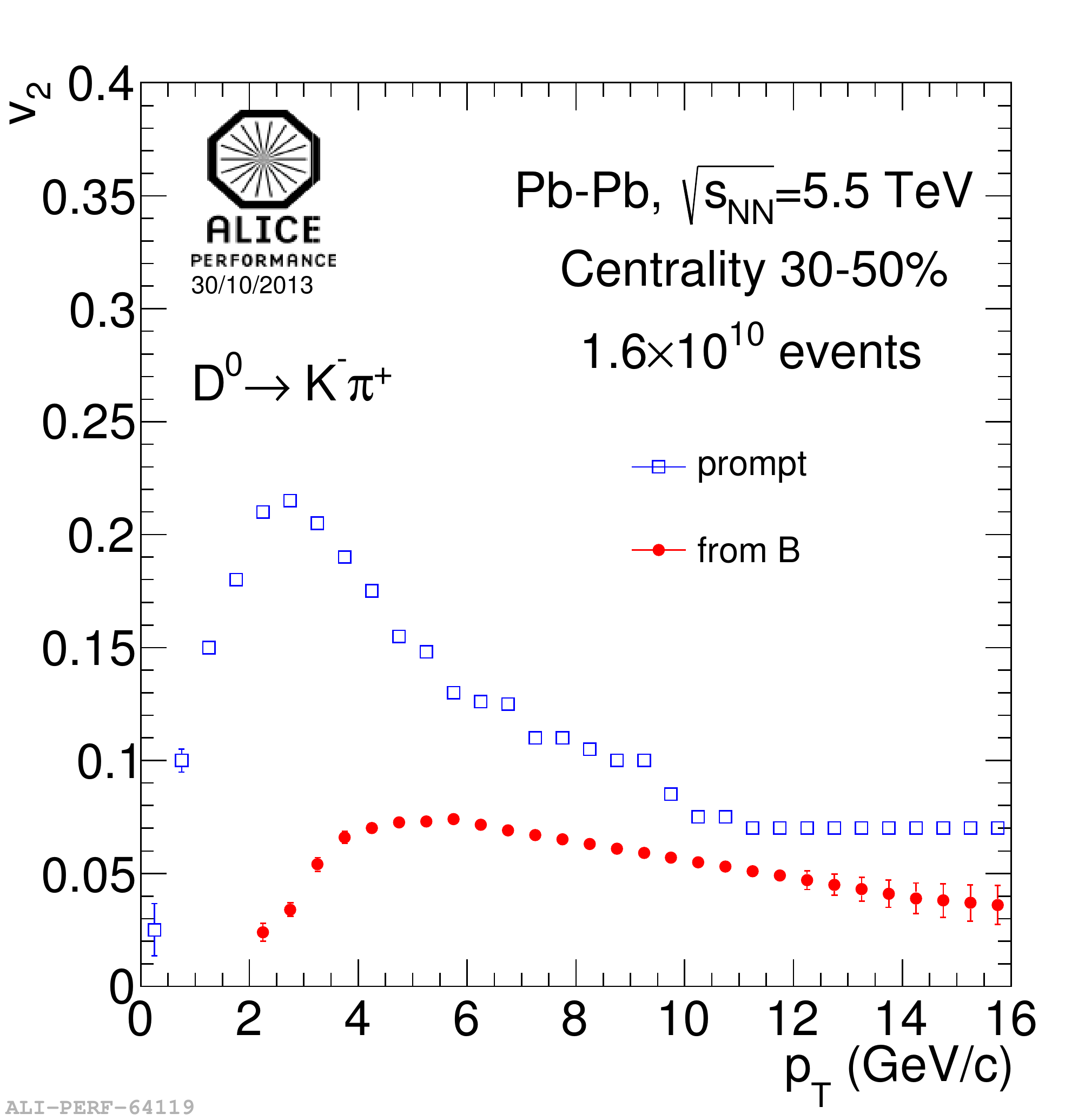} 
  \label{fig:Perf:v2_beauty}
  \caption{Estimated  errors on $v_2$ extraction versus \pt of prompt and non-prompt D$^0$ in \pb collisions~\cite{Abelevetal:2014dna}. }
\end{figure}
After LS2, charm and beauty elliptic flow will be accessible down to zero \pt with an excellent precision (see Figure~\ref{fig:Perf:v2_beauty}). 
The simultaneous measurements of nuclear modification factor and elliptic flow for several hadrons will be a powerful tool to constrain models and extract the QGP transport coefficients.\\
Charmonium is a primary observable of QGP formation.
The first ALICE measurements  suggests that  (re)generation mechanism plays a significant role in low-\pt ($< 3$~GeV/$c$) \jpsi production at LHC energies.
Both the statistical hadronization and kinetic transport model predictions do explain these first observations. 
The measurement of low-\pt $\psi$(2S)/$\psi$ ratio is expected to be sensitive  the dynamics of the $c\bar{c}$ interaction in the medium and therefore to discriminate between kinetic recombination of $c\bar{c}$ pair and statistical hadronization models.
The \RunOne measurement of $\psi$(2S) is statistically limited and due to a very small S/B ratio the $\psi$(2S)  extraction is impossible in central \pb collisions with the present apparatus~\cite{Adam:2015isa}.
High statistics after \LSTwo combined with the new Muon Forward Tracker will provide enough accuracy to discriminate models as shown by left panel of Figure~\ref{fig:Perf:jpsi}.
The \jpsi produced by the recombination of  $c\bar{c}$  pairs in later stages of the collisions would inherit the elliptic flow of the charm quarks in the QGP.
 In this respect, like for the open heavy-flavour particles, the measurement of quarkonium elliptic flow is especially promising to complement the measurements of yields and nuclear modification factors. 
\RunOne provided hint for a non-zero flow of inclusive \jpsi~\cite{ALICE:2013xna}.
As shown on the right panel of Figure~\ref{fig:Perf:jpsi}, after \LSTwo the measurement precision will improve by an order of magnitude.
In addition, MFT will allow the measurement for prompt and non-prompt \jpsi separately  to carried out.
Also, \jpsi $v_2$ measurement will be available at both forward and mid-rapidity.

 \section{CONCLUSION}
Major upgrades of the ALICE experiment are under way to be installed during \LSTwo. 
They will allow the ALICE experiment to be operated at 50~kHz \pb collisions rate, with an improved tracking  performance.
ALICE offers a unique physics programme at LHC, mainly focused on heavy-flavour and charmonium measurements at low \pt, but will also cover the other physics topics not developed here such as the low-mass dileptons, jets, photons or nuclei and anti-nuclei production studies. 
Finally, studies for a forward calorimeter, named FoCal, are on-going.
The goal would be to install a high-granularity calorimeter  at forward rapidity focused on low $x_B$ saturation physics studies.
The installation window of such apparatus would be during LS3.

%

\nocite{*}
\bibliographystyle{aipnum-cp}%
\bibliography{LHCP2015}%

\end{document}